\documentclass[a4paper]{jpconf}
\usepackage{graphicx}
\def \bea{\begin{eqnarray}}
\def \eea{\end{eqnarray}}

\begin{document}
\rightline{EFI 06-25}
\rightline{hep-ph/0612332}
\vskip -1in

\title{Heavy Quark Spectroscopy -- Theory Overview\footnote{Presented at
Second Meeting of APS Topical Group on Hadron Physics, Nashville, TN,
October 22-24, 2006}}

\author{Jonathan L. Rosner}

\address{Enrico Fermi Institute and Department of Physics\\
University of Chicago, 5640 S. Ellis Avenue, Chicago, IL 60637 USA}

\ead{rosner@hep.uchicago.edu}

\begin{abstract}
Some recent discoveries in the spectroscopy of hadrons containing heavy quarks,
and some of their theoretical interpretations, are reviewed.
\end{abstract}

\section{Introduction}
The spectroscopy of states containing heavy quarks $Q$ has undergone a great
renaissance in recent years, providing an exceptional window into
tests of QCD.  Quarkonium systems $Q \bar Q$ are amenable to perturbative
descriptions of their decays.  One can study $Q \bar q$ and $Qqq$ hadrons ($q$
= light quark $u,d,s$) in which the heavy quarks play the role of ``nuclei,''
expanding observables in inverse powers of $m_Q$.  Many heavy-quark hadrons
have masses and couplings strongly affected by nearby thresholds, as has been
known for many years in the physics of atoms and nuclei \cite{Fano,%
Wigner:1948,Feshbach:1958nx}.  Hadron spectra often are crucial in separating
electroweak physics from strong-interaction effects.  More broadly, QCD may not
be the only instance of important non-perturbative effects.  Understanding how
such effects are manifested in hadrons may help prepare us for surprises at the
CERN Large Hadron Collider (LHC).  Finally, at the quark and lepton level there
exists an intricate level structure and a set of transitions calling for
fundamental understanding; spectroscopic methods may help.

We begin this brief review by outlining some theoretical spectroscopic methods.
We then discuss charmed and beauty hadrons, heavy quarkonium ($c \bar c$, $b
\bar b$), and future prospects.

\section{Theoretical methods}

At large distances when the QCD coupling constant becomes too large to permit
the use of perturbation theory, one can place quark and gluon degrees of
freedom on a space-time lattice.  An accurate description of the heavy
quarkonium spectrum then can be obtained once one takes account of degrees of
freedom associated with the pair production of light ($u,d,s$) quarks
\cite{Davies:2003ik}.

Perturbative QCD was applied to charmonium shortly after the discovery of
asymptotic freedom \cite{Appelquist:1974zd}.  It describes $c \bar c$ decays
reasonably well and does better for $b \bar b$ decays, where relativistic
corrections are smaller \cite{Kwong:1987ak,Bethke:2006ac}.

At low energies neither lattice nor perturbative methods are appropriate for
multiparticle systems.  Older techniques of chiral dynamics,
unitarity, and crossing symmetry provide valuable insights for describing
dynamics of mesons up to the GeV scale and baryons somewhat higher.

Hadrons with one charmed or beauty quark can be regarded as ``atoms'' of QCD,
with the light-quark and gluonic degrees of freedom playing the role of the
electron(s) and the heavy quark playing the role of the nucleus.  Properties
of these systems tend to be very simple under the interchange $c
\leftrightarrow b$, in the manner of isotope effects in nuclei.  This {\it
heavy quark symmetry} has led to a number of successful mass and coupling
relations.

For charmonium, and more quantitatively for bottomonium, states
may be described as bound by a potential whose short-distance behavior is
approximately Coulombic and whose long-distance behavior is linear to account
for quark confinement.  Such potential models, often supplemented by
relativistic and/or coupled-channel corrections, provide approximate
descriptions for masses, leptonic partial widths, and hyperfine and
fine-structure splittings.

One can reproduce the spectrum of hadrons containing the
$u,~d$, and $s$ quarks with a model based on additive quark masses
$m_i$ and hyperfine interactions proportional to $\langle \sigma_i \cdot
\sigma_j /(m_i m_j) \rangle$.  The masses of these ``constituent'' quarks are
due in large part to their interaction with the surrounding gluon field.
Correlations between quarks (``diquarks'') also may be important in such
descriptions.  Because of its increased coupling strength at long
distances, QCD leads to the formation of condensates, including non-zero
expectation values of color singlet quark-antiquark pairs and gluonic
configurations such as instantons.  A systematic attempt to cope with the
effect of these condensates on hadron spectroscopy relies on QCD sum rules.

Various phenomenological methods exist for treating resonance decays.  These
mostly rely either on the notion that a single quark in a resonance undergoes a
transition such as pion or photon emission, or the creation of a $q \bar q$
pair corresponding to the breaking of a flux tube connecting constituents of
the resonance.

As this review emphasizes the variety of new experimental data on heavy-quark
spectroscopy seeking theoretical explanation, it will concentrate on schemes
such as lattice and perturbative QCD which have had the greatest predictive
power.  More extensive discussions and references for approaches mentioned
above may be found in Refs.\ \cite{Rosner:2006jz,Eichten:2007}.  Some details
on experimental charm and charmonium results are given in the review of Ref.\
\cite{Lesiak:2006fb}.

\section{Charmed states}

\begin{figure}
\mbox{\includegraphics[width=0.49\textwidth]{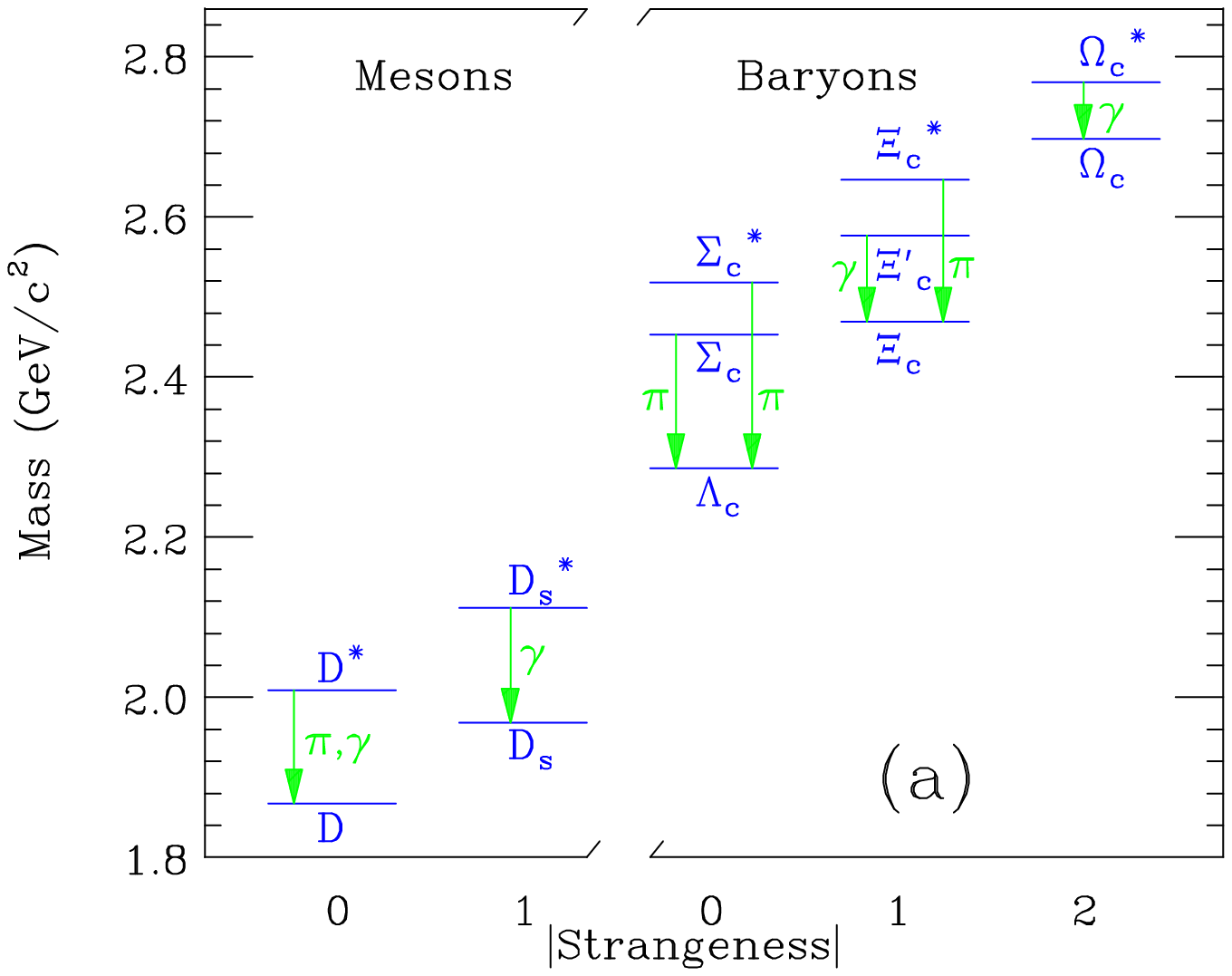}
      \includegraphics[width=0.50\textwidth]{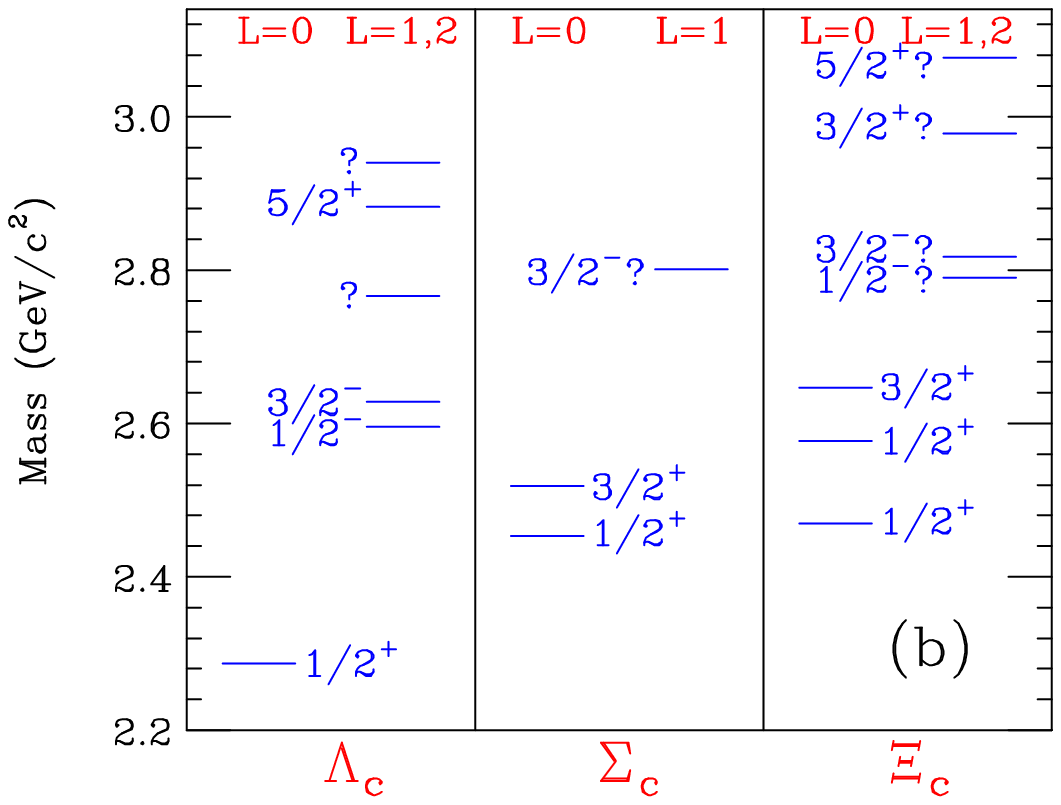}}
\caption{(a) Mesons and baryons containing a single charmed quark in an S-wave.
(b) Charmed baryons and their low-$L$ orbital excitations.
\label{fig:charm}}
\end{figure}

A summary of mesons and baryons containing one charmed quark in an S-wave is
shown in Fig.\ \ref{fig:charm}(a).  The most recent addition is
the $\Omega^*_c$ \cite{Aubert:2006je}, a $css$ candidate for the predicted
$J=3/2$ partner of the $\Omega^- = sss$.  It lies $70.8 \pm 1.0 \pm 1.1$ MeV
above the $\Omega_c$, in agreement with predictions \cite{Rosner:1995yu}.

\subsection{Charmed baryons}

Orbitally-excited charmed baryon levels are plotted along with those of the
lowest $L=0$ states in Fig.\ \ref{fig:charm}(b).  The first excitations of the
$\Lambda_c$ and $\Xi_c$ scale well from the first $\Lambda$ excitations
$\Lambda(1405,1/2^-)$ and $\Lambda(1520,3/2^-)$.  They have the same cost in
$\Delta L$ (about 300 MeV), and their $L \cdot S$ splittings scale as $1/m_s$
or $1/m_c$.  Higher $\Lambda_c$ states may correspond to excitation of a
spin-zero $[ud]$ pair to $S=L=1$, leading to many allowed $J^P$ values up to
$5/2^-$.  In $\Sigma_c$ the light-quark pair has $S=1$; adding $L=1$ allows
$J^P \le 5/2^-$.  States with higher $L$ may be narrower as a result of
increased barrier factors affecting their decays, but genuine spin-parity
analyses would be very valuable.  Some recent results:

(1) The $\Lambda_c(2880)$, first seen in the $\Lambda_c^+ \pi^- \pi^+$ mode
\cite{Artuso:2000xy} and confirmed in the $D^0 p$ mode by BaBar
\cite{Aubert:2006sp}, has been shown to have likely $J^P = 5/2^+$
\cite{Abe:2006rz}.

(2) The highest $\Lambda_c$ was seen by BaBar in the decay mode $D^0 p$
\cite{Aubert:2006sp}.  The Belle Collaboration has seen evidence for its
decay to $\Sigma_c(2455) \pi$ \cite{Abe:2006rz}.

(3) An excited $\Sigma_c$ candidate has been seen decaying to $\Lambda_c \pi^+$,
with mass about 510 MeV above $M(\Lambda_c)$ \cite{Mizuk:2004yu}.  Its $J^P$
shown in Fig.\ \ref{fig:charm}(b) is a guess, using ideas of \cite{SW}, and is
consistent with the assignment proposed in \cite{Lesiak:2006fb} based on the
prediction of \cite{Copley:1979wj}.

(4) The highest $\Xi_c$ levels were reported by the Belle Collaboration in
Ref.\ \cite{Chistov:2006zj}, and confirmed by BaBar \cite{Aubert:2006uw}, both
in the $\Lambda_c^+ K^- \pi^+$ channel.  Their masses suggest $L^P=2^+$.

\begin{figure}
\begin{center}
\includegraphics[width=0.75\textwidth]{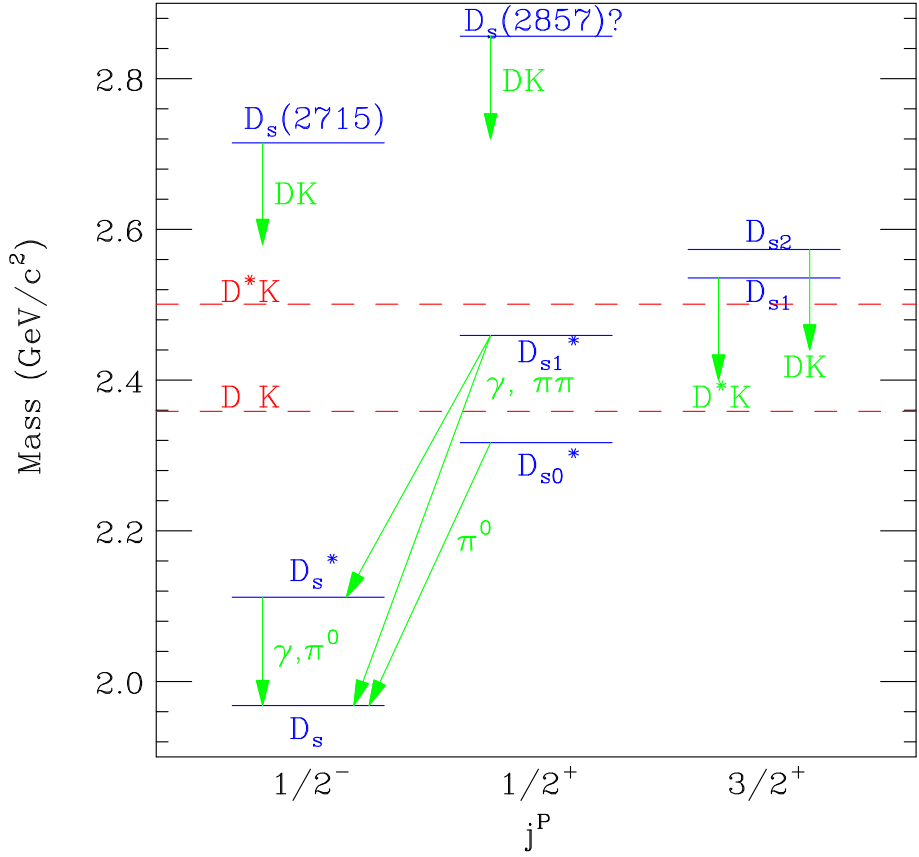}
\end{center}
\caption{Charmed-strange mesons with $L=0$ (negative-parity), $L=1$
(positive-parity), and candidate for state with $L=2$ (positive parity).  Here
$j^P$ denotes the total light-quark
spin + orbital angular momentum and the parity $P$.
\label{fig:ds}}
\end{figure}

\subsection{Excited $D_{sJ}$ and $D$ states}

Excited $D_{sJ}$ states are depicted in Fig.\ \ref{fig:ds}.
The lowest $J^P = 0^+$ and $1^+$ $c \bar s$ states turned out much lighter than
most expectations.  If as heavy as the already-seen $c \bar s$ $L=1$ states,
$D_{s1}(2536)$ [$J^P = 1^+$] and $D_{s2}(2573)$ [$J^P = 2^+$]), they would
have been able to decay to $D \bar K$ (the $0^+$ state) and $D^* \bar K$ (the
$1^+$ state).  Instead a narrow $D_s(2317) \equiv D_{s0}^*$ decaying to $\pi^0
D_s$ and a narrow $D_s(2460) \equiv D_{s1}^*$ decaying to $\pi^0 D_s^*$ were
seen \cite{Aubert:2003fg}.  Their low
masses allow isospin-violating and electromagnetic decays of $D_{s0}^*$ and
$D_{s1}^*$ to be observable.  The decays $D_s(2460)
\to D_s \gamma$ and $D_s(2460) \to D_s \pi^+ \pi^-$ also have been seen
\cite{Marsiske}, and the absolute branching ratios
${\cal B}(D_{s1}^* \to \pi^0 D_s^*) = (0.56 \pm 0.13 \pm 0.09)\%,$
${\cal B}(D_{s1}^* \to \gamma D_s) = (0.16 \pm 0.04 \pm 0.03)\%,$
${\cal B}(D_{s1}^* \to \pi^+ \pi^- D_s^*) = (0.04 \pm 0.01)\%$
measured.

The selection rules in decays of these states show that their $J^P$ values
are consistent with $0^+$ and $1^+$.  Low masses are predicted
\cite{Bardeen:2003kt} if these states are viewed as chiral-symmetry
parity-doublets of the $D_s(0^-)$ and $D^*_s(1^-)$ $c \bar s$ ground states.
The splitting from the ground states is 350 MeV in each case.  Alternatively,
one can view these particles as bound states of $D^{(*)}K$ (the binding energy
in each case would be 41 MeV), or as $c \bar s$ states with masses lowered by
coupling to $D^{(*)}K$ channels \cite{vanBeveren:2003kd,Close:2004ip}.  In
either framework, light-quark degrees of freedom appear to be important in
getting the $D_{s0}^*$ and $D_{s1}^*$ masses right.

A candidate for the first radial excitation of the $D_s^*(2112)$ has been
seen by Belle in $B^+ \to \bar D^0 D^0 K^+$ decays \cite{Abe:2006xm} in the
$M(D^0 K^+)$ spectrum.  Its mass and width are $(2715 \pm 11^{+11}_{-14})$ and
$(115 \pm 20^{+36}_{-32})$ MeV/$c^2$.  Its spin-parity is $J^P = 1^-$.  It lies
$(603^{+16}_{-18})$ MeV/$c^2$ above the ground state, in between the
$2^3S_1$--$1^3S_1$ spacings of $(681 \pm 20)$ MeV/$c^2$ for $s \bar s$
and 589 MeV/$c^2$ for $c \bar c$ \cite{PDG}. This is as expected in a potential
interpolating between $c \bar c$ and $b \bar b$ states \cite{Martin:1980rm},
and as predicted in Ref.\ \cite{Godfrey:1985xj}.  (Ref.\ \cite{Zhang:2006yj}
prefers to identify this state as the lowest $^3D_1$ $c \bar s$ level.)

A higher-lying $c \bar s$ state \cite{Aubert:2006mh} is seen by BaBar decaying
to $D^0 K^+$ and $D^+ K_S$, so it must have natural spin-parity $0^+$, $1^-$,
$2^+, \ldots$.  Its mass and width are $(2856.6 \pm 1.5 \pm 5.0)$ and $(48 \pm
7 \pm 10)$ MeV/$c^2$.  It has been interpreted as a radial excitation of the
$0^+$ state $D_{s0}(2317)$ \cite{vanBeveren:2006st,Close:2006gr}, shown in
Fig.\ \ref{fig:ds}, or a $3^-(^3D_3)$ state \cite{Colangelo:2006rq}.  The same
experiment also sees a broad peak of marginal significance with $M=(2688 \pm 4
\pm 3)$ MeV/$c^2$, $\Gamma = (112 \pm 7 \pm 36)$ MeV/$c^2$.

In contrast to the lightest $0^+$, $1^+$ charmed-strange states, which are
too light to decay to $D K$ or $D^* K$, the lightest $0^+$, $1^+$
charmed-nonstrange candidates appear to be heavy enough to decay to $D \pi$
or $D^* \pi$, and thus are expected to be broad.  Heavy quark symmetry
predicts the existence of a $0^+$, $1^+$ pair with light-quark total angular
momentum and parity $j^P = 1/2^+$ decaying to $D \pi$ or $D^* \pi$,
respectively, via an S-wave.  A $1^+$, $2^+$ pair with $j^P = 3/2^+$, decaying
primarily via a D-wave to $D^* \pi$ or both $D \pi$ and $D^* \pi$,
respectively, is represented by states at $2422.3 \pm 1.3$ MeV/$c^2$ and
$2461.1 \pm 1.6$ MeV/$c^2$ \cite{PDG}.  As for the $j^P = 1/2^+$ candidates,
CLEO \cite{Anderson:1999wn} and Belle \cite{Abe:2003zm} find a broad $1^+$
state in the range 2420--2460 MeV/$c^2$, while Belle and FOCUS
\cite{Abe:2003zm,Link:2003bd,Colangelo:2006aa} find broad $0^+$ candidates
near 2300 and 2400 MeV/$c^2$, respectively.

\subsection{Charmed meson decay constants}

CLEO's value $f_{D^+}=(222.6 \pm 16.7^{+2.8}_{-3.4})$ MeV \cite{Artuso:2005ym}
is consistent with a lattice prediction \cite{Aubin:2005ar}
of $(201 \pm 3 \pm 17)$ MeV.  The accuracy of the previous world average
\cite{PDG} $f_{D_s} = (267 \pm 33)$ MeV has been improved by a BaBar value
$f_{D_s} = 283 \pm 17 \pm 7 \pm 14$ MeV \cite{Aubert:2006sd} and a new
CLEO value $f_{D_s} = 280.1 \pm 11.6 \pm 6.0$ MeV \cite{Stone06}.  The latter,
when combined with CLEO's $f_D$, leads to $f_{D_s}/f_D = 1.26 \pm 0.11 \pm
0.03$.  A lattice prediction for $f_{D_s}$ \cite{Aubin:2005ar} is $f_{D_s} =
249 \pm 3 \pm 16$ MeV, leading to $f_{D_s}/f_D = 1.24 \pm 0.01 \pm 0.07$.
One expects $f_{B_s}/f_B \simeq f_{D_s}/f_D$ so better measurements
of $f_{D_s}$ and $f_D$ by CLEO will help validate lattice calculations and
provide input for interpreting $B_s$ mixing.  A desirable error on $f_{B_s}/f_B
\simeq f_{D_s}/f_D$ is $\le 5\%$ for useful determination of CKM element ratio
$|V_{td}/V_{ts}|$, needing errors $\le 10$ MeV on $f_{D_s}$ and $f_D$.
The ratio $|V_{td}/V_{ts}| = 0.2060 \pm 0.0007~({\rm exp})^{+0.0081}_{-0.0080}
~({\rm theor})$ is implied by a recent CDF result on $B_s$--$\overline{B}_s$
mixing \cite{Abulencia:2006ze} combined with
$B$--$\overline{B}$ mixing and $\xi \equiv (f_{B_s} \sqrt{B_{B_s}}/f_B
\sqrt{B_B}) = 1.21^{+0.047}_{-0.035}$ from the lattice \cite{Okamoto}.
A simple quark model scaling argument anticipated $f_{D_s}/f_D \simeq
f_{B_s}/f_B \simeq \sqrt{m_s/m_d} \simeq 1.25$ \cite{Rosner90}.

\begin{figure}
\includegraphics[width=0.9\textwidth]{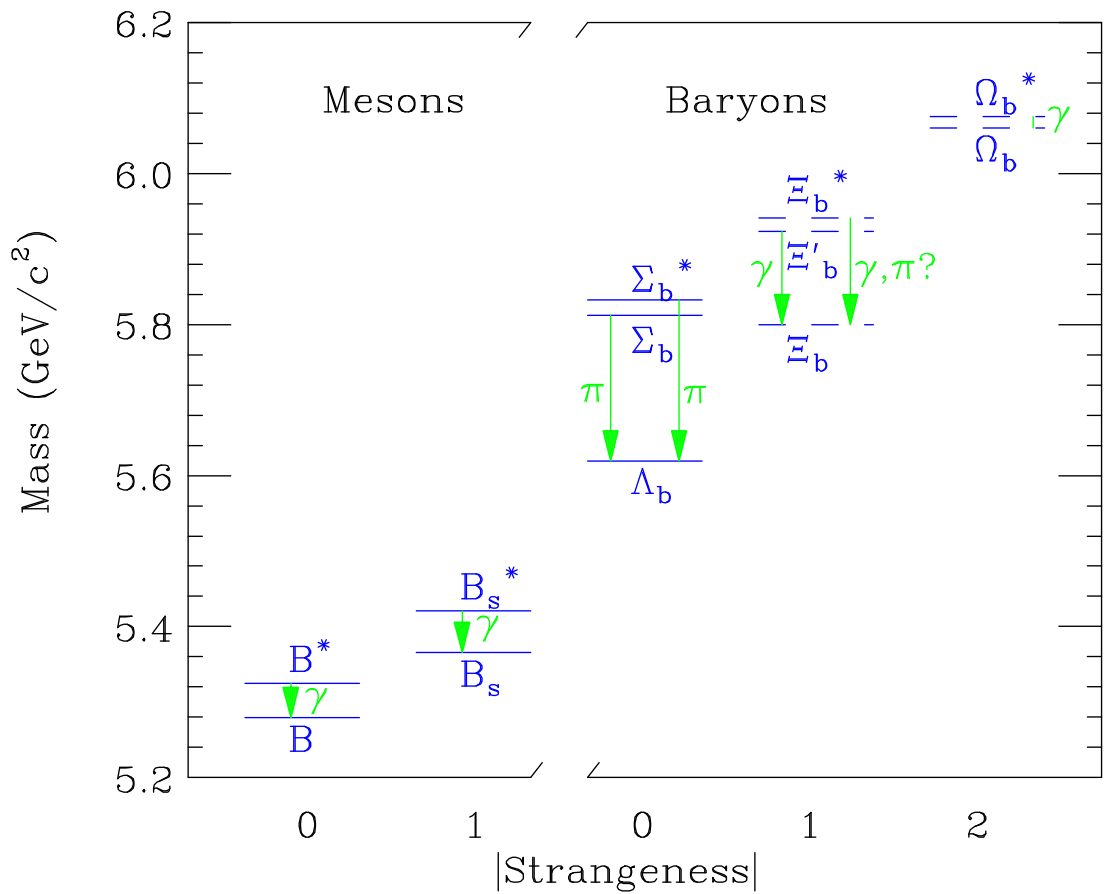}
\caption{S-wave hadrons containing a single beauty quark.  Dashed lines denote
predicted levels not yet observed.
\label{fig:beauty}}
\end{figure}

\section{Beauty hadrons}

The spectrum of ground-state hadrons containing a single $b$ quark is shown in
Fig.\ \ref{fig:beauty}.  The CDF Collaboration has published measurements of
the $B_s$ and $\Lambda_b$ masses and the $B_s$--$B^0$ and $\Lambda_b$--$B^0$
mass differences which are of better precision than the current world averages
\cite{Acosta:2005mq}.  With 1 fb$^{-1}$ CDF now has evidence for the
long-sought $\Sigma_b$ and $\Sigma^*_b$ states very near the masses predicted
from the corresponding charmed baryons using heavy quark symmetry.  (See
\cite{Rosner:2006yk} for some references.) The analysis of Ref.\ \cite{CDFsigb}
studies the spectra of $\Lambda_b \pi^\pm$ states, finding peaks at the values
of $Q^{(*)\pm} \equiv M(\Sigma^{(*)\pm}) - M(\pi^\pm) - M(\Lambda_b)$ shown in
Table \ref{tab:sigb}.  These may be combined with the CDF value $M(\Lambda_b) =
5619.7 \pm 1.7 \pm 1.7$ MeV \cite{Acosta:2005mq} to obtain masses of the
$\Sigma_b^{(*)\pm}$ states.  Here $Q$ and $Q^*$ denote the averages of $Q^\pm$
and $Q^{*\pm}$, respectively.  In this analysis it was assumed that $Q^{*+} -
Q^{*-} = Q^+ - Q^-$.  This assumption was examined in Ref.\
\cite{Rosner:2006yk} and found to be valid to a fraction of an MeV/$c^2$.

\begin{table}
\caption{Values of $Q^{(*)\pm} \equiv M(\Sigma_b^{(*)\pm}) - M(\pi^\pm) -
M(\Lambda_b)$ and $M(\Sigma^{(*)\pm})$ reported by the CDF Collaboration
\cite{CDFsigb}.
\label{tab:sigb}}
\begin{center}
\begin{tabular}{c c} \br
Quantity & Value (MeV) \\ \mr
$Q^+$ & $48.4^{+2.0}_{-2.3} \pm 0.1$ \\
$Q^-$ & $55.9 \pm 1.0 \pm 1.0$ \\
$Q^* - Q$ & $21.3^{+2.0+0.4}_{-1.9-0.2}$ \\ \br
\end{tabular}
\end{center}
\end{table}

A new CDF value for the $\Lambda_b$ lifetime, $\tau(\Lambda_b) = (1.593
^{+0.083}_{-0.078} \pm 0.033)$ ps, was reported recently
\cite{Abulencia:2006dr}.  Whereas
the previous world average of $\tau(\Lambda_b)$ was about 0.8 that of $B^0$,
below theoretical predictions, the new CDF value substantially increases the
world average to a value $\tau(\Lambda_b) = (1.410 \pm 0.054)$ ps which is
$0.923 \pm 0.036$ that of $B^0$ and quite comfortable with theory.

The CDF Collaboration has identified events of the form $B_c \to J/\psi
\pi^\pm$, allowing a precise determination of the
mass: $M$=(6276.5$\pm$4.0$\pm$2.7) MeV/$c^2$ \cite{Aoki:2006}.  This is in
reasonable accord with the latest lattice prediction of
6304$\pm$12$^{+18}_{-0}$ MeV \cite{Allison:2004be}.

The long-awaited $B_s$--$\overline{B}_s$ mixing has finally been observed
\cite{Abulencia:2006ze,D0mix}. The CDF value, $\Delta m_s = 17.77\pm0.10\pm
0.07$ ps$^{-1}$, constrains $f_{B_s}$ and $|V_{td}/V_{ts}|$, as mentioned
earlier.

The Belle Collaboration has observed the decay $B \to \tau \nu_\tau$
\cite{Btaunu}, leading to $f_B |V_{ub}| = (10.1^{+1.6+1.1}_{-1.4-1.3})
\times 10^{-4}$ GeV.  When combined with the value $|V_{ub}| = (4.39 \pm 0.33)
\times 10^{-3}$ \cite{HFAG}, this leads to $f_B = (229^{+36+30}_{-31-34})$
MeV.  A recent lattice estimate \cite{Gray:2005ad} is $f_B =
(216\pm22)$ MeV.

\begin{figure}
\includegraphics[width=0.98\textwidth]{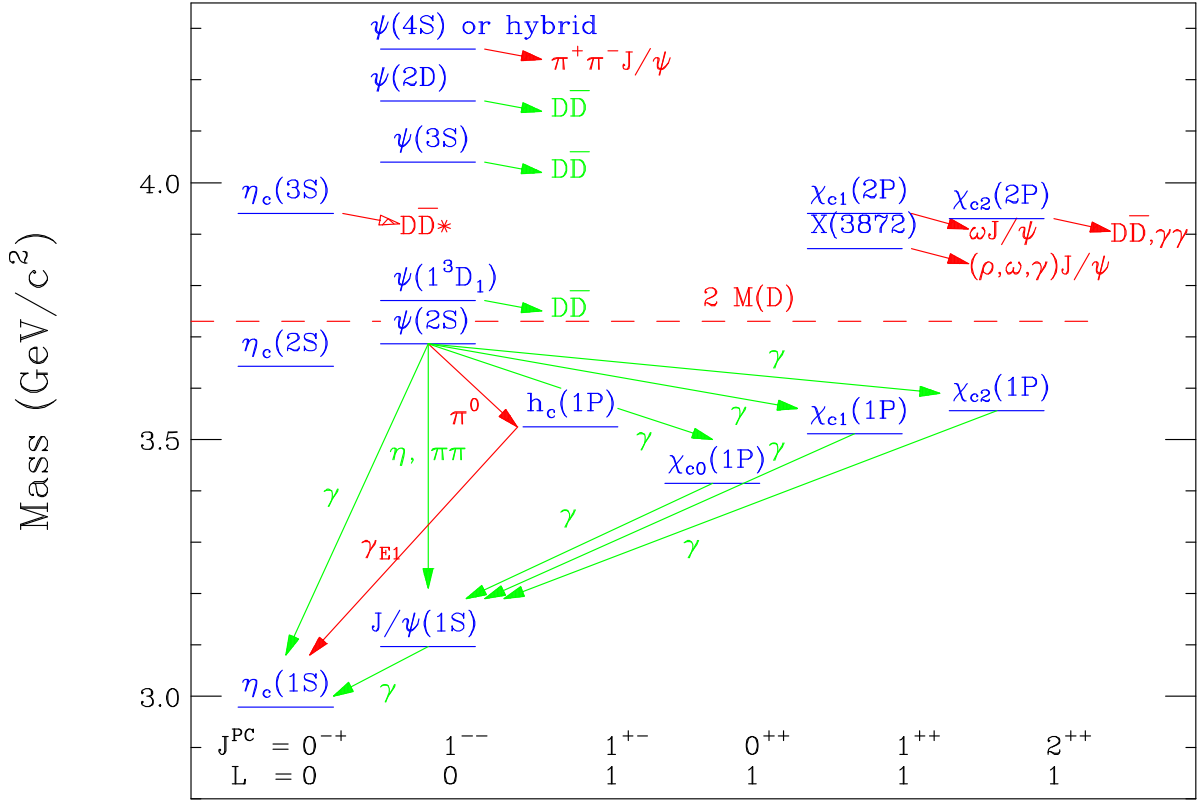}
\caption{Charmonium states including levels above charm threshold.
\label{fig:charmon}}
\end{figure}

\section{Charmonium}

Remarkable progress has been made in the spectroscopy of charmonium states
above charm threshold in the past few years.  Fig.\ \ref{fig:charmon}
summarizes the levels (some of whose assignments are tentative).  Even though
such states can decay to charmed pairs (with the possible exception of
$X(3872)$, which may be just below $D \bar D_1$ threshold), other decay modes
are being seen.  We now discuss some aspects of the recent discoveries.

\subsection{Observation of $h_c$}

The $h_c(1^1P_1)$ state of charmonium has been observed by CLEO
\cite{Rosner:2005ry,Rubin:2005px} via $\psi(2S) \to \pi^0 h_c$ with $h_c
\to \gamma \eta_c$.  Hyperfine splittings test the spin-dependence and spatial
behavior of the $Q \bar Q$ force.  While these are $M(J/\psi) - M(\eta_c)
\simeq 115$ MeV for 1S and $M[\psi'] - M(\eta'_c) \simeq $49 MeV for 2S levels,
P-wave splittings should be less than a few MeV since the potential is
proportional to $\delta^3(\vec{r})$ for a Coulomb-like $c \bar c$ interaction.
Lattice QCD \cite{latt} and relativistic potential \cite{Ebert:2002pp}
calculations confirm this expectation.  One expects $M(h_c) \equiv M(1^1P_1)
\simeq \langle M(^3P_J) \rangle = 3525.36 \pm 0.06$ MeV.  Earlier $h_c$
sightings \cite{Rosner:2005ry,Rubin:2005px} based on $\bar p p$ production in
the direct channel, include a few events seen in CERN ISR Experiment R704; a
state decaying to $\pi^0 J/\psi$ reported by Fermilab E760 but
not confirmed by Fermilab E835; and a state at $3525.8 \pm 0.2 \pm 0.2$ MeV,
decaying to $\gamma \eta_c$ with $\eta_c \to \gamma \gamma$, reported by
E835 with about a dozen candidate events \cite{Andreotti:2005vu}.

In the CLEO data, both exclusive and inclusive analyses see a signal near
$\langle M(^3P_J) \rangle$.  The exclusive analysis reconstructs $\eta_c$ in 7
decay modes and sees a signal of $17.5 \pm 4.5$ events above background.  The
mass and product branching ratio for the two transitions are $M(h_c) = (3523.6
\pm 0.9 \pm 0.5)$ MeV; ${\cal B}_1(\psi' \to \pi^0 h_c) {\cal B}_2(h_c \to
\gamma \eta_c) = (5.3 \pm 1.5 \pm 1.0) \times 10^{-4}$.  Two inclusive analyses
with no $\eta_c$ reconstruction yield $M(h_c) = (3524.9 \pm 0.7 \pm 0.4)$ MeV,
${\cal B}_1 {\cal B}_2 = (3.5 \pm 1.0 \pm 0.7) \times 10^{-4}$.  Combining
exclusive and inclusive results yields $M(h_c) = (3524.4 \pm 0.6 \pm 0.4)$ MeV,
${\cal B}_1 {\cal B}_2 = (4.0 \pm 0.8 \pm 0.7) \times 10^{-4}$.  The $h_c$ mass
is $(1.0 \pm 0.6 \pm 0.4)$ MeV below $\langle M(^3P_J) \rangle$, barely
consistent with the (nonrelativistic) bound \cite{Stubbe:1991qw} $M(h_c) \ge
\langle M(^3P_J) \rangle$ and indicating little P-wave hyperfine splitting in
charmonium.  The value of ${\cal B}_1 {\cal B}_2$ agrees with theoretical
estimates of $(10^{-3} \cdot 0.4)$.

\subsection{$\psi''(3770)$}

The $\psi''(3770)$ is a potential ``charm factory'' for present and future $e^+
e^-$ experiments.  At one time $\sigma(e^+ e^- \to \psi'')$ seemed larger than
$\sigma(e^+ e^- \to \psi'' \to D \bar D)$, implying significant non-$D \bar D$
decays of $\psi''$ \cite{Rosner:2004wy}. A new CLEO measurement \cite{CLDDbar},
$\sigma(\psi'') = (6.38 \pm 0.08 ^{+0.41}_{-0.30})$ nb, appears very close to
the CLEO value $\sigma(D \bar D) = (6.39\pm0.10^{+0.17}_{-0.08})$ nb
\cite{He:2005bs}, leaving little room for non-$D \bar D$ decays.  (BES analyses
\cite{BESsig} do not exclude a 10--20\% non-$D \bar D$ component.)

Some branching ratios for $\psi'' \to X J/\psi$ \cite{Adam:2005mr} are
${\cal B}(\psi'' \to \pi^+ \pi^- J/\psi) =(0.189\pm0.020\pm0.020)\%$,
${\cal B}(\psi'' \to \pi^0 \pi^0 J/\psi) =(0.080\pm0.025\pm0.016)\%$,
${\cal B}(\psi'' \to \eta J/\psi) = (0.087\pm0.033\pm0.022)\%$, and
${\cal B}(\psi'' \to \pi^0 J/\psi) < 0.028\%$.
The value of ${\cal B}[\psi''(3770) \to \pi^+ \pi^- J/\psi]$ found by CLEO is a
bit above 1/2 that reported by BES \cite{Bai:2003hv}.
These account for less than 1/2\% of the total $\psi''$ decays.

CLEO has reported results on $\psi'' \to \gamma \chi_{cJ}$ partial
widths, based on the exclusive process $\psi'' \to \gamma \chi_{c1,2} \to
\gamma \gamma J/\psi \to \gamma \gamma \ell^+ \ell^-$ \cite{Coan:2005} and
reconstruction of exclusive $\chi_{cJ}$ decays \cite{Briere:2006ff}.  The
results are shown in Table \ref{tab:psipprad}, implying
$\sum_J{\cal B}(\psi'' \to \gamma \chi_{cJ}) = {\cal O}$(1\%).

\begin{table}[b]
\caption{CLEO results on radiative decays $\psi'' \to \gamma \chi_{cJ}$.
Theoretical predictions of \cite{Eichten:2004uh} are (a) without and
(b) with coupled-channel effects; (c) shows predictions of
\cite{Rosner:2004wy}.
\label{tab:psipprad}}
\medskip
\begin{center}
\begin{tabular}{ccccc} \br
Mode & \multicolumn{3}{c}{Predicted (keV)} & CLEO \\
     & (a) & (b) & (c) & \cite{Briere:2006ff} \\ \mr
$\gamma \chi_{c2}$ & 3.2 & 3.9 & 24$\pm$4 & $<21$ \\
$\gamma \chi_{c1}$ & 183 & 59 & $73\pm9$ & $75\pm18$ \\
$\gamma \chi_{c0}$ & 254 & 225 & 523$\pm$12 & $172\pm30$ \\ \br
\end{tabular}
\end{center}
\end{table}

Both CLEO and BES \cite{LP123}, in searching for enhanced light-hadron modes,
find only that the $\rho \pi$ mode, suppressed in $\psi(2S)$ decays, also is
{\it suppressed} in $\psi''$ decays.  Several other searches for $\psi''(3770)
\to ({\rm light~ hadrons})$, including VP, $K_L K_S$, and multi-body final
states have been performed.  Two CLEO analyses \cite{Adams:2005ks,Huang:2005fx}
find no evidence for any light-hadron $\psi''$ mode except $\phi \eta$ above
expectations from continuum production.

\subsection{The $X(3872)$}

Many charmonium states above $D \bar D$ threshold have been seen recently
\cite{GodfreyFPCP,Swanson}.  The $X(3872)$, discovered by Belle in $B$ decays
\cite{Choi:2003ue} and confirmed by BaBar \cite{Aubert:2004ns} and in hadronic
production \cite{Acosta:2003zx}, decays predominantly into $J/\psi \pi^+
\pi^-$.  Since it lies well above $D \bar D$ threshold but is narrower than
experimental resolution (a few MeV), unnatural $J^P = 0^-,1^+, 2^-$ is favored.
It has many features in common with an S-wave bound state of $(D^0 \bar D^{*0}
+ \bar D^0 D^{*0})/ \sqrt{2} \sim c \bar c u \bar u$ with $J^{PC} = 1^{++}$
\cite{Close:2003sg}.  The simultaneous decay of $X(3872)$ to $\rho J/\psi$ and
$\omega J/\psi$ with roughly equal branching ratios is a consequence of this
``molecular'' assignment.

Analysis of angular distributions \cite{Rosner:2004ac} in $X \to \rho J/\psi,
\omega J/\psi$ favors the $1^{++}$ assignment \cite{Abe:2005iy} (see also
\cite{Marsiske,Swanson}).  An analysis by the CDF Collaboration
\cite{Abulencia:2006ma} finds equally good fits of decay angular distributions
to $J^{PC} = 1^{++}$ and $2^{-+}$.  The latter is disfavored by Belle's
observation \cite{Gokhroo:2006bt} of $X \to D^0 \bar D^0 \pi^0$, which would
require at least two units of relative orbital angular momentum in the
three-body state, very near threshold.  Observation of the $\gamma J/\psi$ mode
($\sim 14\%$ of $J/\psi \pi^+ \pi^-$) \cite{Abe:2005ix} confirms the $C=+$
assignment and suggests a $c \bar c$ admixture in the wave function.  BaBar
\cite{Bapipipsi} finds ${\cal B}[X(3872) \to \pi^+ \pi^- J/\psi] > 0.042$ at
90\% c.l.

\subsection{Charmonium between 3.9 and 4.0 GeV/$c^2$}

Belle has reported a candidate for a $2^3P_2(\chi'_{c2})$ state in $\gamma
\gamma$ collisions \cite{Abe:2005bp}, decaying to $D \bar D$.
The angular distribution of $D \bar D$ pairs is
consistent with $\sin^4 \theta^*$ as expected for a state with $J=2, \lambda =
\pm2$.  It has $M = 3929 \pm 5 \pm 2$ MeV, $\Gamma = 29 \pm 10 \pm 3$ MeV, and
$\Gamma_{ee} {\cal B}(D \bar D) = 0.18 \pm 0.06 \pm 0.03$ eV, all reasonable
for a $\chi'_{c2}$ state.

A charmonium state $X(3938)$ is produced recoiling against $J/\psi$ in $e^+ e^-
\to J/\psi + X$ \cite{Abe:2005hd} and is seen decaying to $D \bar D^*$ +
c.c.  Since all lower-mass states observed in this recoil process have $J=0$
[these are the $\eta_c(1S), \chi_{c0}$ and $\eta'_c(2S)$], it is tempting to
identify this state with $\eta_c(3S)$ (not $\chi'_{c0}$, which would decay to
$D \bar D$).

The $\omega J/\psi$ final state in $B \to K \omega J/\psi$ shows a peak above
threshold at $M(\omega J/\psi) \simeq 3940$ MeV \cite{Abe:2004zs}.  This could
be a candidate for one or more excited P-wave charmonium states, likely the
$\chi'_{c1,2}(2^3P_{1,2})$.  The corresponding $b \bar b$ states $\chi'_{b1,2}$
have been seen to decay to $\omega \Upsilon(1S)$ \cite{Severini:2003qw}.

\subsection{The $Y(4260)$}

BaBar has reported a state $Y(4260)$ produced in the radiative return
reaction $e^+ e^- \to \gamma \pi^+ \pi^- J/\psi$ and seen in the $\pi^+ \pi^-
J/\psi$ spectrum \cite{Aubert:2005rm}.  Its mass is consistent with being a
$4S$ level \cite{Llanes-Estrada:2005vf} since it lies about 230 MeV above the
$3S$ candidate (to be compared with a similar $4S$-$3S$ spacing in the
$\Upsilon$ system).  The level spacings of charmonium and bottomonium would be
identical if the interquark potential were $V(r) \sim {\rm log}(r)$, which may
be viewed as an interpolation between the short-distance $\sim -1/r$ and
long-distance $\sim r$ behavior expected in QCD \cite{Quigg:1977dd}.  Other
interpretations of $Y(4260)$ include a $c s \bar c \bar s$ state
\cite{Maiani:2005pe} and a hybrid $c \bar c g$ state \cite{Zhu:2005hp},
for which it lies in the expected mass range.

The CLEO Collaboration has confirmed the $Y(4260)$, both in a direct scan
\cite{Coan:2006rv} and in radiative return \cite{He:2006kg}.  Signals are seen
for $Y(4260) \to \pi^+ \pi^- J/\psi$ 11$\sigma$), $\pi^0 \pi^0 J/\psi$
(5.1$\sigma$), and $K^+ K^- J/\psi$ (3.7$\sigma$).  There are also weak signals
for $\psi(4160) \to \pi^+ \pi^- J/\psi$ (3.6$\sigma$) and $\pi^0 \pi^0 J/\psi$
(2.6$\sigma$), consistent with the $Y(4260)$ tail, and for $\psi(4040) \to
\pi^+ \pi^- J/\psi$ (3.3$\sigma$).  Both CLEO and Belle \cite{Abe:2006hf} see
the state at slightly higher mass than BaBar.

The hybrid interpretation of $Y(4260)$ deserves further attention.  One
consequence is a predicted decay to $D \bar D_1 +$ c.c., where $D_1$
is a P-wave $c \bar q$ pair.  Now, $D \bar D_1$ threshold is 4287 MeV/$c^2$
if we consider the lightest $D_1$ to be the state noted in Ref.\ \cite{PDG}
at 2422 MeV/$c^2$.  In this case the $Y(4260)$ would be a $D \bar D_1 +$ c.c.
{\it bound state}.  It would decay to $D \pi \bar D^*$, where the $D$ and $\pi$
are not in a $D^*$. The dip in $R_{e^+ e^-}$ lies just below $D \bar D_1$
threshold, which may be the first S-wave meson pair accessible in
$c \bar c$ fragmentation \cite{Close:2005iz}.  The $D^* \bar D_0^*$ mode
could also be either another decay channel of $Y(4260)$ or represent a
separate resonance with slightly greater mass and width.

\section{Bottomonium}

\begin{figure}
\begin{center}
\includegraphics[height=0.4\textheight]{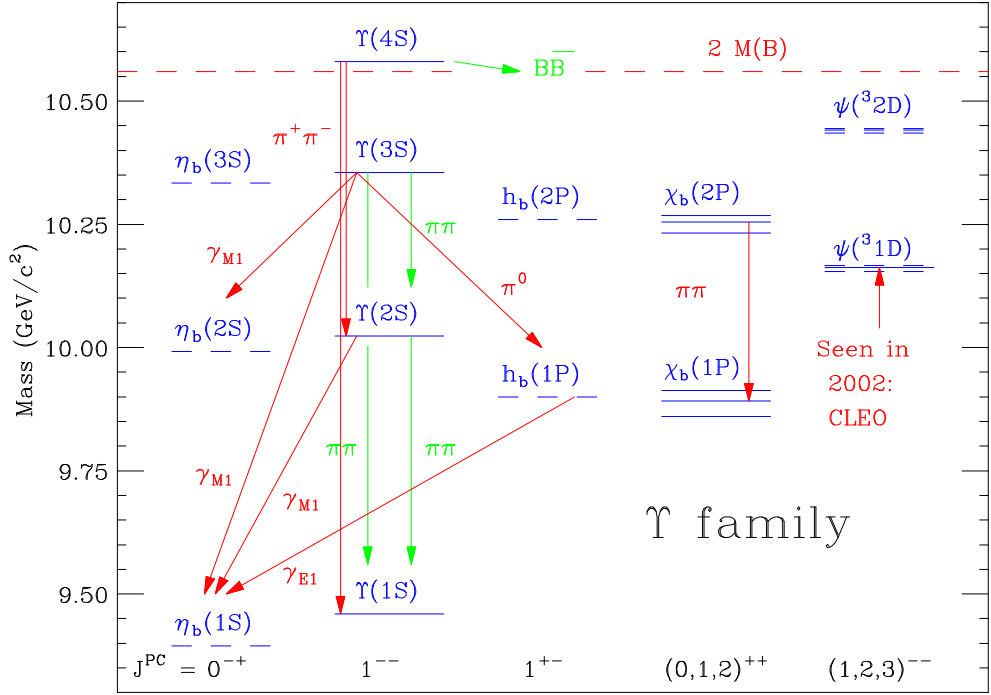}
\end{center}
\caption{$b \bar b$ levels and some decays.  Electric dipole (E1) transitions
$S \leftrightarrow P \leftrightarrow D$ are not shown.
\label{fig:ups}}
\end{figure}

Some properties and decays of the $\Upsilon$ ($b \bar b$) levels are summarized
in Fig.\ \ref{fig:ups}.  Masses are in agreement with unquenched lattice QCD
calculations \cite{Lepage}.  Direct photons have been
observed in 1S, 2S, and 3S decays, implying estimates of the strong
fine-structure constant consistent with others \cite{Besson:2005jv}.  The
transitions $\chi_b(2P) \to \pi \pi \chi_b(1P)$ have been seen
\cite{Cawlfield:2005ra,Tati}.  BaBar has measured the partial widths
$\Gamma[\Upsilon(4S) \to \pi^+ \pi^- \Upsilon(1S)] = 1.8 \pm 0.4$ keV and
$\Gamma[\Upsilon(4S) \to \pi^+ \pi^- \Upsilon(2S)] = 2.7 \pm 0.8$ keV
\cite{Aubert:2006bm}, while Belle
has seen $\Upsilon(4S) \to \pi^+ \pi^- \Upsilon(1S)$, with a branching ratio
${\cal B} = (1.1 \pm 0.2 \pm 0.4) \times 10^{-4}$ \cite{BeUps}.

\subsection{Remeasurements by the CLEO Collaboration}

New values of ${\cal B}[\Upsilon(1S,2S,3S) \to \mu^+ \mu^-] = (2.49 \pm 0.02
\pm 0.07,~2.03\pm0.03\pm0.08,~2.39\pm0.07\pm0.10)\%$ \cite{Adams:2004xa},
when combined with new measurements $\Gamma_{ee}(1S,2S,3S) = (1.252\pm0.004
\pm0.019,~0.581\pm0.004\pm0.009,~0.413\pm0.004\pm0.006)$ keV and previous
data, imply total widths \cite{PDG} $\Gamma_{\rm tot}(1S,2S,3S) = (54.02\pm
1.25,~31.98\pm2.63,~20.32\pm1.85)$ keV.  The values of $\Gamma_{\rm tot}
(2S,3S)$ are significantly below previous world averages \cite{PDG04}, leading
to changes in comparisons of predicted and observed transition rates.  As one
example, the study of $\Upsilon(2S,3S) \to \gamma X$ decays
\cite{Artuso:2004fp} has provided new branching ratios for E1 transitions
to $\chi_{bJ}(1P),~\chi'_{bJ}(2P)$ states.  These may be combined with the
new total widths to obtain updated partial decay widths [Table \ref{tab:E1},
line (a)], which may be compared with one set of non-relativistic
predictions \cite{KR} [line (b)].  The suppression of transitions to $J=0$
states by 10--20\% with respect to non-relativistic expectations agrees
with relativistic predictions \cite{rel}.  The partial width for $\Upsilon(3S)
\to \gamma 1^3P_0$ is found to be $61 \pm 23$ eV, about nine times the
highly-suppressed value predicted in Ref.\ \cite{KR}.  That prediction is
very sensitive to details of wave functions; the discrepancy indicates
the importance of relativistic distortions.

\begin{table}
\caption{Comparison of observed (a) and predicted (b) partial widths
for $2S \to 1 P_J$ and $3S \to 2 P_J$ transitions in $b \bar b$ systems.
\label{tab:E1}}
\medskip
\begin{center}
\begin{tabular}{|c|c c c|c c c|} \br
 & \multicolumn{3}{c|}{$\Gamma$ (keV), $2S \to 1P_J$ transitions}
 & \multicolumn{3}{c|}{$\Gamma$ (keV), $3S \to 2P_J$ transitions} \\
 & $J=0$ & $J=1$ & $J=2$ & $J=0$ & $J=1$ & $J=2$ \\ \mr
(a) & 1.20$\pm$0.18 & 2.22$\pm$0.23 & 2.32$\pm$0.23 &
 1.38$\pm$0.19 & 2.95$\pm$0.30 & 3.21$\pm$0.33 \\
(b) &     1.39       &     2.18      &     2.14      &
     1.65      &     2.52       &     2.78       \\ \br
\end{tabular}
\end{center}
\end{table}

\subsection{$b \bar b$ spin singlets}

Decays of the $\Upsilon(1S,2S,3S)$ states are potential sources of information
on $b \bar b$ spin-singlets, but none has been seen yet.  One expects
1S, 2S, and 3S hyperfine splittings to be approximately 60, 30, 20 MeV/$c^2$,
respectively \cite{Godfrey:2001eb}.  The lowest P-wave singlet state (``$h_b$'')
is expected to be near $\langle M(1^3P_J) \rangle \simeq 9900$ MeV/$c^2$
\cite{Godfrey:2002rp}.

Several searches have been performed or are under way in 1S, 2S, and 3S CLEO
data.  One can search for the allowed M1 transition in $\Upsilon(1S) \to \gamma
\eta_b(1S)$ by reconstructing exclusive final states in $\eta_b(1S)$ decays
and dispensing with the soft photon, which is likely to be swallowed up in
background.  Final states are likely to be of high multiplicity.

One can search for higher-energy but suppressed M1 photons in $\Upsilon(n'S)
\to \gamma \eta_b(nS)\\(n \ne n')$ decays.  These searches already exclude many
models. The strongest upper limit obtained is for $n'=3$, $n=1$: ${\cal B} \le
4.3 \times 10^{-4}$ (90\% c.l.).  $\eta_b$ searches using sequential processes
$\Upsilon(3S) \to \pi^0 h_b(1^1P_1) \to \pi^0 \gamma \eta_b(1S)$ and
$\Upsilon(3S) \to \gamma \chi'_{b0} \to \gamma \eta \eta_b(1S)$ (the latter
suggested in Ref.\ \cite{Voloshin:2004hs}) are being conducted but there are no
results yet.  Additional searches for $h_b$ involve the transition
$\Upsilon(3S) \to \pi^+ \pi^- h_b$ [for which a typical experimental upper
bound based on earlier CLEO data \cite{Brock:1990pj} is
${\cal O}(10^{-3}$)], with a possible $h_b \to \gamma \eta_b$ transition
expected to have a 40\% branching ratio \cite{Godfrey:2002rp}.

\section{Future prospects}

CLEO and BES-III will make new
contributions to heavy quark spectroscopy.  CLEO will focus on center-of-mass
energies 3770 and 4170 MeV, to obtain about 750 pb$^{-1}$ at each energy. Goals
include the best possible determination of $f_D$ and $f_{D_s}$, measurements
of form factors for semileptonic $D$ and $D_s$ decays which will provide
incisive tests for lattice gauge theories, and measurement of the CKM factors
$V_{cd}$ and $V_{cs}$ with unprecedented precision.  CLEO collected over 26
million $\psi(2S)$ (about 8 times the current sample) this past summer and
looks forward to fruitful analyses of these data.  CLEO-c running will
end at the end of March 2008; BES-III will take over, and PANDA (a proposed
detector in Germany) is anticipated to begin running in 2014.

Belle has taken 2.9 fb$^{-1}$ of data at $\Upsilon(3S)$.  They have been
concerned primarily with ``invisible'' decays of the $\Upsilon(1S)$ [also the
subject of a CLEO search], tagged via $\Upsilon(3S) \to \pi^+ \pi^-
\Upsilon(1S)$.  This sample is also potentially valuable for spectroscopy.
CLEO has (1.1, 1.2, 1.2) fb$^{-1}$ at 1S, 2S, 3S.
Both BaBar and Belle have shown interest in hadron spectroscopy; they are
well-positioned to study it.  There have been useful contributions from CDF
and D0 as well.

Hadron spectroscopy is providing both long-awaited
states like $h_c$ (whose mass and production rate confirm theories of quark
confinement and isospin-violating $\pi^0$-emission transitions) and surprises
like low-lying P-wave $D_s$ mesons, X(3872), X(3940), Y(3940), Z(3940) and
Y(4260).  Decays of the $\psi''(3770)$ have been important in confirming its
interpretation as a D-wave $c \bar c$ state with some S-wave admixture.  
We are continuing to learn about properties of QCD in the strong-coupling
regime through evidence for molecules, 3S, 2P, 4S or hybrid charmonium, and
interesting decays of states above flavor threshold.

QCD may not be the last strongly coupled theory with which we have to deal.
Understanding the mystery of electroweak symmetry breaking or the very
structure of quarks and leptons may require related techniques.
These insights are coming to us in general from experiments at
the frontier of intensity and detector capabilities rather than energy,
and illustrate the importance of a diverse approach to the fundamental
structure of matter.

\ack{I thank I. Gorelov, P. Maksimovic, and W. Wester for helpful discussions.
This work was supported in part by the United States Department of
Energy under Grant No.\ DE-FG02-ER40560.}


\medskip

\begin{thebibliography}{99}

\bibitem{Fano} U. Fano, Nuovo Cim.\ {\bf 12}, 154 (1935) [Translation: J. Res.\
Natl. Inst.\ Stand.\ Technol.\ {\bf 110}, 583 (2005)]; Phys.\ Rev.\ {\bf 124},
1866 (1961).

\bibitem{Wigner:1948} E. P. Wigner, Phys.\ Rev.\ {\bf 73}, 1002 (1948).

\bibitem{Feshbach:1958nx} H.~Feshbach, Ann. Phys.\ (N.Y.) {\bf 5}, 357 (1958).

\bibitem{Davies:2003ik} C.~T.~H.~Davies {\it et al.} [HPQCD Collaboration],
  Phys.\ Rev.\ Lett.\ {\bf 92}, 022001 (2004).

\bibitem{Appelquist:1974zd} T.~Appelquist and H.~D.~Politzer,
  Phys.\ Rev.\ Lett.\ {\bf 34}, 43 (1975).

\bibitem{Kwong:1987ak}W.~Kwong, P.~B.~Mackenzie, R.~Rosenfeld and J.~L.~Rosner,
  Phys.\ Rev.\ D {\bf 37}, 3210 (1988).

\bibitem{Bethke:2006ac} S.~Bethke, hep-ex/0606035; S. Kluth, hep-ex/0609020,
presented at ICHEP 06 (Moscow, Russia, 2006), Proceedings edited by A. N.
Sissakian and G. A. Kozlov, to be published by World Scientific, 2007.

\bibitem{Aubert:2006je} B.~Aubert {\it et al.} [BABAR Collaboration],
  arXiv:hep-ex/0608055.

\bibitem{Rosner:1995yu} J.~L.~Rosner, Phys.\ Rev.\ D {\bf 52}, 6461 (1995).

\bibitem{Rosner:2006jz} J.~L.~Rosner, arXiv:hep-ph/0609195, submitted to
J. Phys.\ G.

\bibitem{Eichten:2007} E. Eichten, S. Godfrey, H. Mahlke-Kr\"uger, and J. L.
Rosner, 2007, Enrico Fermi Institute Report No.\ EFI 06-15, in preparation.

\bibitem{Lesiak:2006fb} T.~Lesiak,
  hep-ex/0612042, presented at HQL06, Munich, Oct.\ 16--20, 2006.

\bibitem{Artuso:2000xy} M.~Artuso {\it et al.} [CLEO Collaboration],
  Phys.\ Rev.\ Lett.\ {\bf 86}, 4479 (2001).

\bibitem{Aubert:2006sp} B.~Aubert {\it et al.} [BABAR Collaboration],
  SLAC report SLAC-PUB-11786, hep-ex/0603052.

\bibitem{Abe:2006rz} K. Abe {\it et al.} [Belle Collaboration],
  Belle report BELLE-CONF-0602, hep-ex/0608043, submitted to ICHEP 06,
 {\it op.\ cit.}

\bibitem{Mizuk:2004yu} R.~Mizuk {\it et al.} [Belle Collaboration],
Phys.\ Rev.\ Lett.\ {\bf 94}, 122002 (2005).

\bibitem{SW} F.~Wilczek, hep-ph/0409168, in {\it From fields to strings:
Circumnavigating theoretical physics: Ian Kogan memorial collection},
edited by M. Shifman {\it et al.}, vol. 1 (World Scientific, Singapore,
2005), p.\ 77;  A. Selem, Senior Thesis, M. I. T., 2005 (unpublished);
A.~Selem and F.~Wilczek, hep-ph/0602128, in {\it Proc.\
Ringberg Workshop On New Trends In HERA Physics 2005}, 2--7 October 2005,
Tegernsee, Germany, edited by G. Grindhammer {\it et al.}
(World Scientific, Hackensack, NJ, 2006), p.\ 337.

\bibitem{Copley:1979wj} L.~A.~Copley, N.~Isgur and G.~Karl,
  Phys.\ Rev.\ D {\bf 20}, 768 (1979)
  [Erratum-ibid.\ D {\bf 23}, 817 (1981)].

\bibitem{Chistov:2006zj} R.~Chistov {\it et al.} [BELLE Collaboration],
Phys.\ Rev.\ Lett.\  {\bf 97}, 162001 (2006).

\bibitem{Aubert:2006uw} B.~Aubert {\it et al.} [BABAR Collaboration],
  SLAC-PUB-11980, BABAR-CONF-06-01, hep-ex/0607042, submitted to ICHEP 06,
{\it op.\ cit.}

\bibitem{Aubert:2003fg} B.~Aubert {\it et al.}  [BaBar Collaboration],
Phys.\ Rev.\ Lett.\ {\bf 90}, 242001 (2003); D.~Besson {\it et al.} [CLEO
Collaboration], Phys.\ Rev.\ D {\bf 68}, 032002 (2003); K.~Abe {\it et al.}
[Belle Collaboration], Phys.\ Rev.\ Lett.\ {\bf 92}, 012002 (2004).

\bibitem{Marsiske} H. Marsiske, at Flavor Physics and CP Violation Conference,
(FPCP 2006) Vancouver, BC, April, 2006, Proceedings, p.\ 012; B.~Aubert {\it et
al.} [BaBar Collaboration], Phys.\ Rev.\ D {\bf 74}, 031103 (2006); S. J. Gowdy
[for the BaBar Collaboration], at Moriond 2006 (QCD and Hadronic
Interactions at High Energy), hep-ex/0605086.

\bibitem{Bardeen:2003kt} W.~A.~Bardeen, E.~J.~Eichten and C.~T.~Hill,
Phys.\ Rev.\ D {\bf 68}, 054024 (2003), and refs.\ therein.

\bibitem{vanBeveren:2003kd} E.~van Beveren and G.~Rupp,
Phys.\ Rev.\ Lett.\ {\bf 91}, 012003 (2003);
Eur.\ Phys.\ J.\ C {\bf 32}, 493 (2004).

\bibitem{Close:2004ip} F.~E.~Close,
  Int.\ J.\ Mod.\ Phys.\ A {\bf 20}, 5156 (2005).

\bibitem{Abe:2006xm} K. Abe {\it et al.} [Belle Collaboration], Belle report
BELLE-CONF-0643, hep-ex/0608031, submitted to ICHEP 06, {\it op.\ cit.}

\bibitem{PDG} W.-M. Yao {\it et al.} [Particle Data Group], J. Phys.\ G
{\bf 33}, 1 (2006).

\bibitem{Martin:1980rm} A.~Martin, Phys.\ Lett.\ B {\bf 100}, 511 (1981).

\bibitem{Godfrey:1985xj} S.~Godfrey and N.~Isgur,
  Phys.\ Rev.\ D {\bf 32}, 189 (1985).

\bibitem{Zhang:2006yj} B.~Zhang, X.~Liu, W.~Z.~Deng and S.~L.~Zhu,
  hep-ph/0609013.

\bibitem{Aubert:2006mh} B.~Aubert {\it et al.} [BaBar Collaboration],
SLAC report SLAC-PUB-11993, hep-ex/0607082.

\bibitem{vanBeveren:2006st} E.~van Beveren and G.~Rupp,
Phys.\ Rev.\ Lett.\ {\bf 97}, 202001 (2006).

\bibitem{Close:2006gr} F.~E.~Close, C.~E.~Thomas, O.~Lakhina and E.~S.~Swanson,
  hep-ph/0608139.

\bibitem{Colangelo:2006rq} P.~Colangelo, F.~De Fazio and S.~Nicotri,
  Phys.\ Lett.\ B {\bf 642}, 48 (2006).

\bibitem{Anderson:1999wn} S.~Anderson {\it et al.} [CLEO Collaboration],
  Nucl.\ Phys.\ A {\bf 663}, 647 (2000).

\bibitem{Abe:2003zm} K.~Abe {\it et al.} [Belle Collaboration],
  Phys.\ Rev.\ D {\bf 69}, 112002 (2004).

\bibitem{Link:2003bd} J.~M.~Link {\it et al.} [FOCUS Collaboration],
  Phys.\ Lett.\ B {\bf 586}, 11 (2004).

\bibitem{Colangelo:2006aa} P.~Colangelo, F.~De Fazio, R.~Ferrandes and
 S.~Nicotri, hep-ph/0609240.

\bibitem{Artuso:2005ym} M.~Artuso {\it al.} [CLEO Collaboration],
Phys.\ Rev.\ Lett.\ {\bf 95}, 251801 (2005).

\bibitem{Aubin:2005ar}
C.~Aubin {\it et al.}, Phys.\ Rev.\ Lett.\ {\bf 95}, 122002 (2006).

\bibitem{Aubert:2006sd} B.~Aubert {\it et al.} [BABAR Collaboration],
hep-ex/0607094, submitted to Phys.\ Rev.\ Letters.

\bibitem{Stone06} S. Stone [CLEO Collaboration], presented at ICHEP 06,
{\it op.\ cit.},\\
{\tt http://ichep06.jinr.ru/reports/179\_10s2\_18p05\_Stone.pdf}

\bibitem{Abulencia:2006ze} A.~Abulencia {\it et al.} [CDF Collaboration],
hep-ex/0609040, superseding an earlier result:  A. Abulencia {\it et al.} [CDF
Collaboration], Phys.\ Rev.\ Lett.\ {\bf 97}, 062003 (2006).

\bibitem{Okamoto} M.~Okamoto, PoS {\bf LAT2005}, 013 (2006), hep-lat/0510113.

\bibitem{Rosner90} J. L. Rosner, Phys.\ Rev.\ D {\bf 42}, 3732 (1990).

\bibitem{Acosta:2005mq} D.~Acosta {\it et al.} [CDF Collaboration],
  Phys.\ Rev.\ Lett.\ {\bf 96}, 202001 (2006).

\bibitem{Rosner:2006yk} J.~L.~Rosner,
  arXiv:hep-ph/0611207, submitted to Phys.\ Rev.\ D.

\bibitem{CDFsigb} J. Pursley, talk presented on behalf of the CDF Collaboration
at the 11th International Conference on $B$-Physics at Hadron Machines
(Beauty 2006), 25--29 September 2006, University of Oxford, to be published in
Nucl.\ Phys.\ B (Proc.\ Suppl.);
I. Gorelov, presented on behalf of the CDF Collaboration
at the Second Meeting of the APS Topical Group on Hadron Physics (GHP 2006),
22--24 October 2006, Nashville, TN; CDF Collaboration, public web page \\
{\tt http://www-cdf.fnal.gov/physics/new/bottom/060921.blessed-sigmab/}.

\bibitem{Aoki:2006} M. Aoki [for the CDF Collaboration], presented to
Quarkonium Working Group, Brookhaven Natl.\ Lab., June 27--30,
2006, CDF Public Note 8004 (unpublished), updating D.~Acosta {\it et al.},
Phys.\ Rev.\ Lett.\ {\bf 96}, 082002 (2006).

\bibitem{Allison:2004be} I.~F.~Allison, C.~T.~H.~Davies, A.~Gray,
A.~S.~Kronfeld, P.~B.~Mackenzie and J.~N.~Simone [HPQCD Collaboration],
Phys.\ Rev.\ Lett.\  {\bf 94}, 172001 (2005).

\bibitem{D0mix} V. M. Abazov {\it et al.} [D0 Collaboration],
Phys.\ Rev.\ Lett.\ {\bf 97}, 021802 (2006).

\bibitem{Btaunu} Values quoted by K. Ikado {\it et al.},
hep-ex/0604018, hep-ex/0605068, have been corrected for revised efficiency
estimates: Belle report BELLE-CONF-0671, reported by T.
Browder at ICHEP 06, {\it op.\ cit.}

\bibitem{HFAG} Periodic updates may be found at
{\tt http://www.slac.stanford.edu/xorg/hfag/}.

\bibitem{Gray:2005ad} A.~Gray {\it et al.} [HPQCD Collaboration],
  Phys.\ Rev.\ Lett.\ {\bf 95}, 212001 (2005).

\bibitem{Abulencia:2006dr} A.~Abulencia {\it et al.} [CDF Collaboration],
  Fermilab report FERMILAB-PUB-06-321-E, hep-ex/0609021, submitted to Phys.\
  Rev.\ Letters.

\bibitem{Rosner:2005ry} J.~L.~Rosner {\it et al.} [CLEO Collaboration], Phys.\
Rev.\ Lett.\ {\bf 95}, 102003 (2005).

\bibitem{Rubin:2005px} P. Rubin {\it et al.}  [CLEO Collaboration], Phys.\
Rev.\ D {\bf 72}, 092005 (2005).

\bibitem{latt} T. Manke {\it et al.} [CP-PACS Collaboration], Phys.\ Rev.\ D
{\bf 62}, 114508 (2000); M. Okamoto {\it et al.} [CP-PACS Collaboration],
{\it ibid.} {\bf 65}, 095408 (2002).

\bibitem{Ebert:2002pp} D.~Ebert, R.~N.~Faustov and V.~O.~Galkin,
  Phys.\ Rev.\ D {\bf 67}, 014027 (2003), and references therein;
  Mod.\ Phys.\ Lett.\ A {\bf 20}, 1887 (2005).

\bibitem{Andreotti:2005vu} M.~Andreotti {\it et al.} [Fermilab E835
Collab.], Phys.\ Rev.\ D {\bf 72}, 032001 (2005).

\bibitem{Stubbe:1991qw}
J.~Stubbe and A.~Martin, Phys.\ Lett.\ B {\bf 271}, 208 (1991).

\bibitem{Rosner:2004wy} J.~L.~Rosner, Ann.\ Phys.\ (N.Y.) {\bf 319}, 1 (2005).

\bibitem{CLDDbar} D. Besson {\it et al.} [CLEO Collaboration],
Phys.\ Rev.\ Lett.\ {\bf 96}, 092002 (2006).

\bibitem{He:2005bs} Q.~He {\it et al.}  [CLEO Collaboration],
  Phys.\ Rev.\ Lett.\  {\bf 95}, 121801 (2005)
  [Erratum-ibid.\  {\bf 96}, 199903 (2006)].

\bibitem{BESsig} M. Ablikim {\it et al.} [BES Collaboration], Phys.\ Lett.\ B
{\bf 641}, 145 (2006); Phys.\ Rev.\ Lett.\ {\bf 97}, 121801 (2006);
hep-ex/0612056.

\bibitem{Adam:2005mr} N.~E.~Adam [CLEO Collaboration], Phys.\ Rev.\ Lett.\
{\bf 96}, 082004 (2006).

\bibitem{Bai:2003hv} J.~Z.~Bai {\it et al.} [BES Collaboration],
Phys.\ Lett.\ B {\bf 605}, 63 (2005).

\bibitem{Coan:2005} T. E. Coan {\it et al.} [CLEO Collaboration], Phys.\ Rev.\
Lett.\ {\bf 96}, 182002 (2006).

\bibitem{Briere:2006ff} R.~A.~Briere {\it et al.} [CLEO Collaboration],
Phys.\ Rev.\ D {\bf 74}, 031106 (2006).

\bibitem{Eichten:2004uh} E.~J.~Eichten, K.~Lane and C.~Quigg,
  Phys.\ Rev.\ D {\bf 69}, 094019 (2004);
  Phys.\ Rev.\ D {\bf 73}, 014014 (2006)
  [Erratum-ibid.\ D {\bf 73}, 079903 (2006)].

\bibitem{LP123} M.~Ablikim {\it et al.}  [BES Collaboration],
paper no.\ 123, 22nd International Symposium on Lepton-Photon Interactions at
High Energy (LP 2005), Uppsala, Sweden, 30 June - 5 July 2005.

\bibitem{Adams:2005ks} G. S. Adams {\it et al.} [CLEO Collaboration], Phys.\
Rev.\ D {\bf 73}, 012002 (2006).

\bibitem{Huang:2005fx} G.~S.~Huang {\it et al.} [CLEO Collaboration],
  Phys.\ Rev.\ Lett.\ {\bf 96}, 032003 (2006).

\bibitem{GodfreyFPCP} S. Godfrey, invited talk at FPCP 2006, {\it op.\ cit.},
hep-ph/0605152, Proceedings, p.\ 015. 

\bibitem{Swanson} E. Swanson, Conference on the Intersections of Particle and
Nuclear Physics (CIPANP 2006), Rio Grande, Puerto Rico, May 30 -- June 3,
2006, AIP Conference Proceedings Vol.\ 870, edited by T. M. Liss, p.\ 349.

\bibitem{Choi:2003ue} S.~K.~Choi {\it et al.}  [Belle Collaboration],
Phys.\ Rev.\ Lett.\  {\bf 91}, 262001 (2003).

\bibitem{Aubert:2004ns} B.~Aubert {\it et al.} [BaBar Collaboration],
Phys.\ Rev.\ D {\bf 71}, 071103 (2005).

\bibitem{Acosta:2003zx} D.~Acosta {\it et al.} [CDF II Collaboration],
Phys.\ Rev.\ Lett.\  {\bf 93}, 072001 (2004);
V.~M.~Abazov {\it et al.}  [D0 Collaboration],
Phys.\ Rev.\ Lett.\  {\bf 93}, 162002 (2004).

\bibitem{Close:2003sg} F.~E.~Close and P.~R.~Page, Phys.\ Lett.\ B {\bf 578},
119 (2004); N.~A.~Tornqvist, hep-ph/0308277; E.~S.~Swanson, Phys.\ Lett.\ B
{\bf 588}, 189 (2004); {\it ibid.} {\bf 598}, 197 (2004).

\bibitem{Rosner:2004ac} J.~L.~Rosner, Phys.\ Rev.\ D {\bf 70}, 094023 (2004).

\bibitem{Abe:2005iy} K. Abe {\it et al.}, Belle report BELLE-CONF-0541,
hep-ex/0505038, paper no.\ LP-2005-176, LP 2005, {\it op.\ cit.}

\bibitem{Abulencia:2006ma} A. Abulencia {\it et al.} [CDF Collaboration],
  arXiv:hep-ex/0612053.

\bibitem{Gokhroo:2006bt} G.~Gokhroo {\it et al.} [Belle Collaboration],
  Phys.\ Rev.\ Lett.\ {\bf 97}, 162002 (2006).

\bibitem{Abe:2005ix} K. Abe {\it et al.}, Belle report BELLE-CONF-0540,
hep-ex/0505037, paper no.\ LP-2005-175, LP 2005, {\it op.\ cit.}

\bibitem{Bapipipsi} B. Aubert {\it et al.} [BaBar Collaboration], Phys.\ Rev.\
Lett.\ {\bf 96}. 052002 (1996).

\bibitem{Abe:2005bp} K.~Abe {\it et al.} [Belle Collaboration], Phys.\ Rev.\
Lett.\ {\bf 96}. 082003 (2006).

\bibitem{Abe:2005hd} K.~Abe {\it et al.} [Belle Collaboration], hep-ex/0507019,
submitted to Phys.\ Rev.\ Letters; L. Hinz, CIPANP 2006, {\it op.\ cit.}, p.\ 
345.

\bibitem{Abe:2004zs} K.~Abe {\it et al.}  [Belle Collaboration],
Phys.\ Rev.\ Lett.\  {\bf 94}, 182002 (2005).

\bibitem{Severini:2003qw} H.~Severini {\it et al.}  [CLEO Collaboration],
Phys.\ Rev.\ Lett.\  {\bf 92}, 222002 (2004).

\bibitem{Aubert:2005rm} B.~Aubert {\it et al.} [BaBar Collaboration], Phys.\
Rev.\ Lett.\ {\bf 95}, 142001 (2005).

\bibitem{Llanes-Estrada:2005vf} F.~J.~Llanes-Estrada, Phys.\ Rev.\ D {\bf 72},
031503 (2005).

\bibitem{Quigg:1977dd} C.~Quigg and J.~L.~Rosner, Phys.\ Lett.\ B {\bf 71}, 153
(1977).

\bibitem{Maiani:2005pe} L.~Maiani, V.~Riquer, F.~Piccinini and A.~D.~Polosa,
Phys.\ Rev.\ D {\bf 72}, 031502 (2005).

\bibitem{Zhu:2005hp} S.~L.~Zhu, Phys.\ Lett.\ B {\bf 625}, 212 (2005);
E.~Kou and O.~Pene, Phys.\ Lett.\ B {\bf 631}, 164 (2005);
F.~E.~Close and P.~R.~Page, Phys.\ Lett.\ B {\bf 628}, 215 (2005).

\bibitem{Coan:2006rv} T.~E.~Coan {\it et al.}  [CLEO Collaboration],
Phys.\ Rev.\ Lett.\  {\bf 96}, 162003 (2006).

\bibitem{He:2006kg} Q.~He  {\it et al.} [CLEO Collaboration],
Phys.\ Rev.\ D {\bf 74}, 091104 (2006).

\bibitem{Abe:2006hf} K.~Abe {\it et al.} [Belle Collaboration],
  arXiv:hep-ex/0612006.

\bibitem{Close:2005iz} F.~E.~Close and P.~R.~Page,
  Phys.\ Lett.\ B {\bf 628}, 215 (2005).

\bibitem{Lepage} G. Peter Lepage, Ann.\ Phys.\ {\bf 315}, 193 (2005).

\bibitem{Besson:2005jv} D.~Besson {\it et al.} [CLEO Collaboration],
Phys.\ Rev.\ D {\bf 74}, 012003 (2006).

\bibitem{Cawlfield:2005ra} C.~Cawlfield {\it et al.} [CLEO Collaboration],
Phys.\ Rev.\ D {\bf 73}, 012003 (2006).

\bibitem{Tati} G. Tatishvili, CIPANP 2006, {\it op.\ cit.}, p.\ 356.

\bibitem{Aubert:2006bm} B.~Aubert {\it et al.} [BABAR Collaboration],
Phys.\ Rev.\ Lett.\  {\bf 96}, 232001 (2006).

\bibitem{BeUps} K. Abe {\it et al.} [Belle Collaboration], hep-ex/0512034,
contributed to LP 2005 ({\it op.\ cit.}), and to EPS International Europhysics
Conference on High Energy Physics (HEP-EPS 2005), Lisbon, Portugal.

\bibitem{Adams:2004xa} G.~S.~Adams {\it et al.}  [CLEO Collaboration],
Phys.\ Rev.\ Lett.\ {\bf 94}, 012001 (2005).

\bibitem{PDG04} S. Eidelman {\it et al.} [Particle Data Group],
Phys.\ Lett.\ B {\bf 592}, 1 (2004).

\bibitem{Artuso:2004fp} M.~Artuso {\it et al.}  [CLEO Collaboration],
Phys.\ Rev.\ Lett.\ {\bf 94}, 032001 (2005).

\bibitem{KR} W. Kwong and J. L. Rosner, Phys.\ Rev.\ D {\bf 38}, 3179 (1988).

\bibitem{rel} P. Moxhay and J. L. Rosner, Phys.\ Rev.\ D {\bf 28}, 1132 (1983);
R. McClary and N. Byers, {\it ibid.} {\bf 28}, 1692 (1983).  See also
T.~Skwarnicki, hep-ex/0505050, at 40th Rencontres de Moriond On
QCD and High Energy Hadronic Interactions, 12-19 Mar 2005, La Thuile, Aosta
Valley, Italy.

\bibitem{Godfrey:2001eb} S.~Godfrey and J.~L.~Rosner, Phys.\ Rev.\ D {\bf 64},
074011 (2001) [Erratum-ibid.\ D {\bf 65}, 039901 (2002)].

\bibitem{Godfrey:2002rp} S.~Godfrey and J.~L.~Rosner, Phys.\ Rev.\ D {\bf 66},
014012 (2002).

\bibitem{Voloshin:2004hs} M.~B.~Voloshin, Mod.\ Phys.\ Lett.\ A {\bf 19}, 2895
(2004).

\bibitem{Brock:1990pj} I.~C.~Brock {\it et al.} [CLEO Collaboration],
Phys.\ Rev.\ D {\bf 43}, 1448 (1991).

\end{thebibliography}
\end{document}